\documentstyle[11pt,cs10,epsf]{article}

\begin{document}

\title{Supersaturation in X-ray emission of clusters stars} 

\author{Sofia Randich\altaffilmark{1,2}}
\altaffiltext{1}{European Southern Observatory, 
Karl-Schwarzschild-Str. 2, D-85748
Garching bei M\"unchen, Germany}
\altaffiltext{2}{Osservatorio Astrofisico di Arcetri, Largo E. Fermi 5, I-50125
Firenze, Italy}

\begin{abstract}
A population of cool dwarfs with extreme rotational velocities
(v$\sin i \ga 100$ km/sec) is
present in young open clusters.
ROSAT observations have shown that these very fast rotators exhibit
a level of X-ray activity a factor of 3--5 below the saturated level
which is usually observed (both in X-rays and other magnetic activity
indicators) for `normally' fast rotators. This phenomenon has been
denominated {\it ``Supersaturation"}.

W UMa contact binaries seem to be supersaturated as well, while the
scatter in the rotation-activity relation for RS CVn and BY Dra binaries
does not allow us to clearly discern whether they exhibit supersaturation
or not. Supersaturation is not seen in H$\alpha$ for $\alpha$ Per 
supersaturated stars.

Two alternative lines of interpretation are discussed.
\end{abstract}


\section{Introduction}

Saturation of magnetic activity is, from an observational point of view,
a well known phenomenon, though 
it represents an open problem within the framework of dynamo theory.

{\it Einstein}, IUE, and, later, HST and ROSAT observations of stars
in open clusters and in the field have shown that 
the ratio of chromospheric and transition region line
fluxes over bolometric
flux ($f_{\rm line}/f_{\rm bol}$), as well as the X-ray to bolometric
luminosity ratio ($\rm L_{\rm X}/\rm L_{\rm bol}$),
generally increases with increasing
rotation, until a saturation plateau is reached (e.g., Vilhu and
Rucinsky, 1983; Vilhu 1984; Simon and Fekel 1987; Simon 1990; 
Stauffer et al. 1994; Randich et al. 1996;
Ayres et al. 1996; Patten and Simon 1996; Stauffer et al. 1997ab).
In other words, saturation indicates that, outside of flaring, 
the radiation emitted
from plasma at temperatures of $\sim 10^4$, $10^5$, and $10^6$-$10^7$ K,
respectively, 
cannot exceed a given fraction of the total stellar flux, even in most
active stars.
For X-rays such a fraction is 1/1000, or $L_{\rm X}/\rm L_{\rm
bol}=10^{-3}$.

Saturation lacks an agreed upon interpretation; 
it is not clear whether it reflects the saturation of dynamo
itself or it is rather due to the
total filling of the star's surface with active regions, as originally 
suggested by Vilhu (1984).
We mention in passing, that saturation in angular momentum loss
needs to be introduced in the models of angular momentum evolution
in order to explain the fast rotators which are observed in young
clusters (e.g., Collier Cameron and Li 1994; Barnes and Sofia 1996;
Krishnamurthi et al. 1997; Bouvier et al. 1997).

\begin{figure}
\plotfiddle{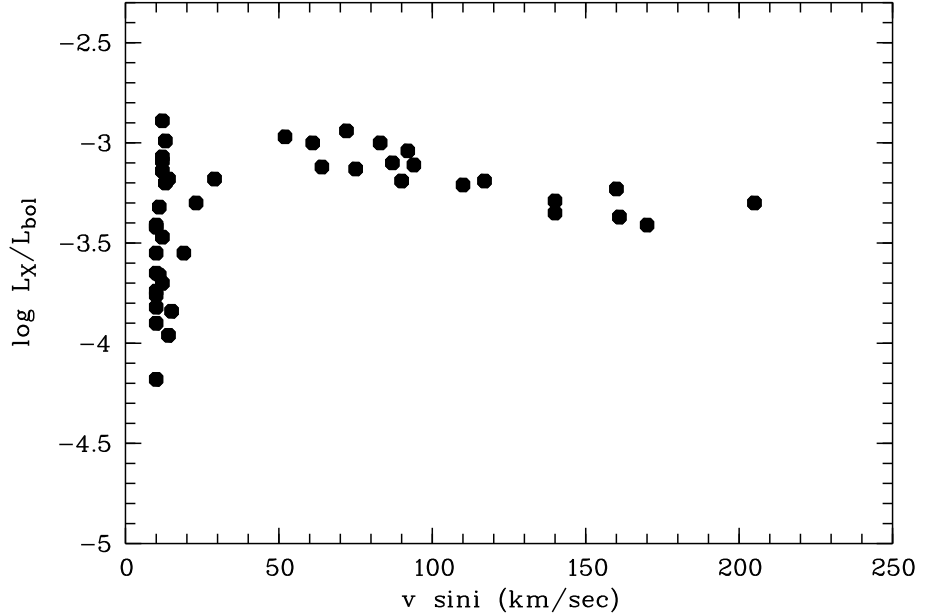}{8cm}{-90}{60}{60}{-220}{320}
\caption{Ratio of X-ray to bolometric luminosity for $\alpha$ Persei
stars. Stars with 0.6 $<$ B--V$_0 <$1.45 are considered.} \label{fig-1}
\end{figure}

The threshold velocity above which saturation is seen is different for
the various activity diagnostics and depends on stellar mass.
If one studies the activity -- rotation relation
considering the Rossby number (the ratio of 
the rotational period to the convective turnover time, $N_{\rm R} = 
\rm P/\tau_{\rm c}$)
instead of rotational velocity or period itself (e.g., Noyes et al 1984;
Simon et al. 1985), 
a critical Rossby number $(N_{\rm R})_{\rm crit.}$ is determined below
which the relation saturates (e.g., Patten and Simon 1996; Stauffer et
al. 1997a). Since $\tau_{\rm c}$ is a function of stellar mass (namely, it
increases with decreasing mass), saturation will start to be seen at
progressively larger periods for stars of lower masses.
Stauffer et al. (1997a), using Noyes et al.'s analytical approximation
for $\tau_{\rm c}$, estimated that saturation should occur at P = 2d
(v(rot) $\sim$ 25 km/sec) for 1 $M_{\odot}$ stars and at P = 4.5d 
(v(rot) $\sim$ 5 km/sec) for 0.4 $M_{\odot}$ stars.
In addition, Stauffer et al (1997b) recently suggested that different
rotational histories, and specifically the presence or not of
a circumstellar disk locking the star during pre-main sequence
contraction, could 
lead to different dynamo-induced magnetic
activity, and eventually to different saturation velocity thresholds.

So much, as far as the low velocity tail of the saturation relation is
concerned. What about the high velocity end, then?
Very young open clusters, like the
50 Myr old $\alpha$ Persei or the 30 Myr old IC 2602 and IC 2391,
have a population of cool ultra fast rotators (UFRs), or G and K-type dwarfs
with projected rotational velocities exceeding 100 km/sec.
ROSAT observations of the IC clusters (Randich et al. 1995; Patten and
Simon 1996) and of $\alpha$ Per (Randich et al. 1996; Prosser et al.
1996) covered many of these UFRs, allowing to study what happens to
the activity-rotation relation for stars with extreme
rotational velocities. 

\index{*$\alpha$ Persei}
\index{*IC 2602}
\index{*IC 2391}
\section{What is ``supersaturation"?}
\begin{figure}
\plotfiddle{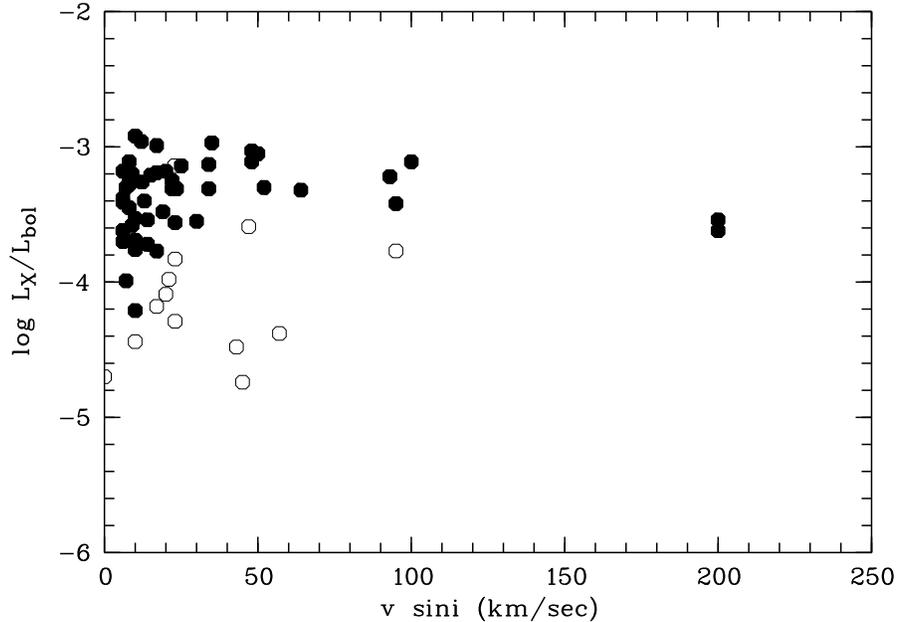}{8cm}{-90}{60}{60}{-220}{320}
\caption{Same as figure ~\ref{fig-1}, but IC 2391 and IC 2602 members are
shown. Filled and open circles denote stars with (V-I)$_{\rm C0} < 0.70$
and 0.4 $<$ (V-I)$_{\rm C0} < $0.70, respectively.} \label{fig-2}
\end{figure}

\begin{figure}
\plotfiddle{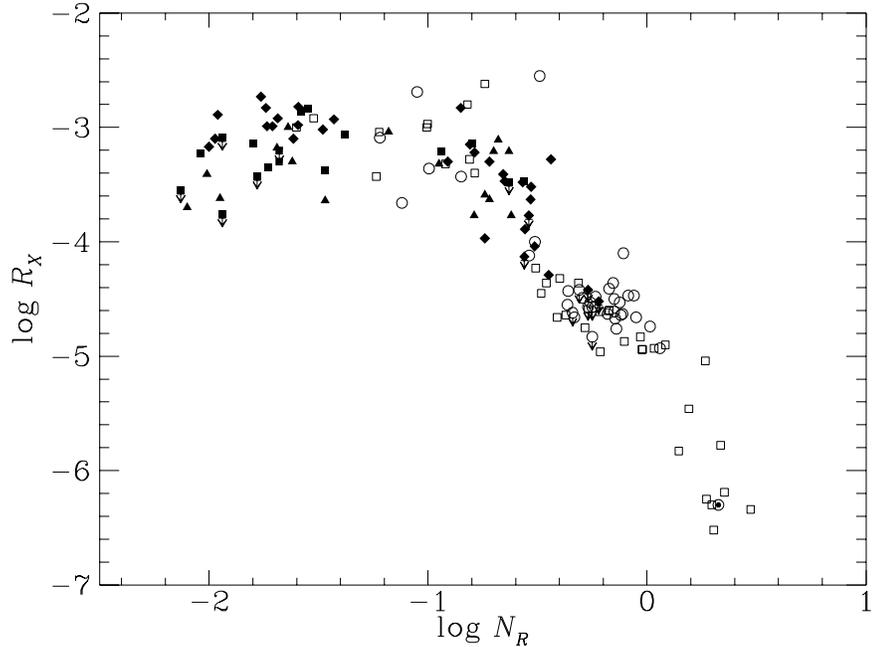}{8cm}{0}{60}{60}{-180}{-100}
\caption{$\log$ L$_{\rm X}$/L$_{\rm bol}$ (R$_{\rm X}$) is plotted vs.
$\log$ Rossby number ($\log N_{\rm R}$) for: IC 2391 (filled triangles),
$\alpha$ Per (filled squares), the Pleiades (filled diamonds), the Hyades
(open circles), and main sequence field stars (open squares). The Sun is
also plotted.} \label{fig-3} 
\end{figure}

In Figure~\ref{fig-1} $\log \rm L_{\rm X}/L_{\rm bol}$ ($\log
R_{\rm X}$) is plotted vs.
v$\sin i$ for $\alpha$ Persei; Figure~\ref{fig-2} is the same, but
IC 2602 and IC 2391 stars are shown,
 whereas Figure~\ref{fig-3}
(from Patten and Simon 1996), shows 
$\log R_{\rm X}$ vs. the logarithm of the Rossby number ($\log N_{\rm R}$)
for stars in the field, and in the IC 2391, $\alpha$ Per, Pleiades, and
Hyades clusters.

\index{*Pleiades}
\index{*Hyades}

The three figures show similar patterns: above a critical velocity
or below a critical
Rossby number, the saturation plateau discussed in the Introduction
is clearly seen (the scatter in the IC 2602/2391 plot for v$\sin i$ between
15 and 30 km/sec reflects the scatter in mass and the fact that
stars with different masses have different saturation thresholds;
see the discussion in Stauffer et al. 1997a). However, most
surprisingly, UFRs (or, more generally, stars with very low Rossby
numbers) do not lie at the saturated level $\log \rm R_{\rm X}=-3$, but
show a decline from it.
This phenomenon has been termed ``{\bf supersaturation}" by Prosser et al.
(1996). 

Supersaturation starts to be seen
for v$\sin i \ga 100$ km/sec, or for Rossby numbers
$\log N_{\rm R}$ between $\sim -1.6$ and $\sim -1.8$.
The decline in $\log \rm R_{\rm X}$, which is as large as a factor of 3--5,
is observed for a significant number of stars and in more than one cluster,
thus I regard as unlikely that it is
due to uncertainties or errors in the estimates of
X-ray luminosities. Moreover, it is also improbable that supersaturation
is `produced' by some effect of rapid rotation on bolometric
luminosity. Whereas very rapid rotation has been shown to modify 
stellar structure, resulting in a change of colors and luminosity (e.g.,
Kraft 1970; Maeder 1971), this effect is at most of the order of 20 \%,
much smaller than the observed decrease in $\log \rm R_{\rm X}$.
I therefore conclude 
that supersaturation indicates a real decrease of X-ray emission
--at least in the ROSAT passband.

\section{Discussion}

Before discussing the possible causes of supersaturation, I wish to
address two questions;
First, is supersaturation 
seen in other classes of stars (specifically, fast rotators) apart
from the ones in young clusters? Second, is 
supersaturation observed for other magnetic activity indicators?

\begin{figure}
\plotfiddle{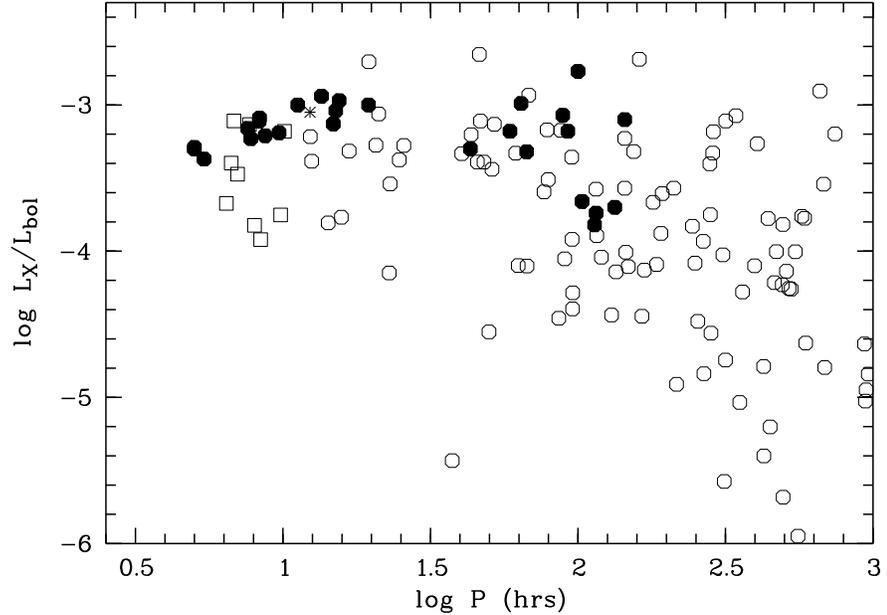}{8cm}{-90}{60}{60}{-230}{320}
\caption{$\log \rm R_{\rm X}$ vs. period for: $\alpha$ Per (filled
circles), RS CVn and By Dra binaries (open circles), W UMa systems
(open squares), and AB Dor (asterisk).} \label{fig-4}
\end{figure}

In Figure ~\ref{fig-4} $\log \rm R_{\rm X}$ vs. $\log$ P is plotted for
$\alpha$ Per stars, 
the sample of BY Draconis and RS Canum
Venaticorum active binaries from Dempsey et al. (1997) and
Dempsey et al. (1993) \footnote{Note
that Dempsey et al. (1997)
do not find any statistically significant differences between the
coronal properties of RS CVn and BY Dra systems and, in particular,
they state that the dependence of X-ray flux on period is the
same for the two groups.}, and for a small sample of W UMa
contact binaries from McGale et al. (1997);
the active star AB Dor is also plotted 
in the figure (asterisk; data from K\"urster et al. 1997).

Fig. ~\ref{fig-4} first shows that,
as noted by Dempsey et al., the distribution of active stars
is characterized by a significant scatter; a few short-period
but very low R$_{\rm X}$ systems are present and, at the same time,
stars at the saturation level, or close to it, are seen already
at longer periods than in $\alpha$ Per. AB Dor, with a period of
12.5 hrs, is at the saturation level.
Most of the very-short
periods binaries, which however have longer periods than $\alpha$
Per supersaturated stars, are indeed below
the saturation level; however,
since the
saturation plateau for RS CVn and BY Dra binaries 
is not as sharp as for the young clusters,
it is difficult to ascertain whether this is a real decline in
$\rm R_{\rm X}$ for the most rapidly spinning binaries. On the other hand, 
as suggested by S. Drake in his question at the Workshop, W UMa binaries seem
to  be `supersaturated', showing a decline from saturation even
larger than UFRs in $\alpha$ Per.

\begin{figure}
\plotfiddle{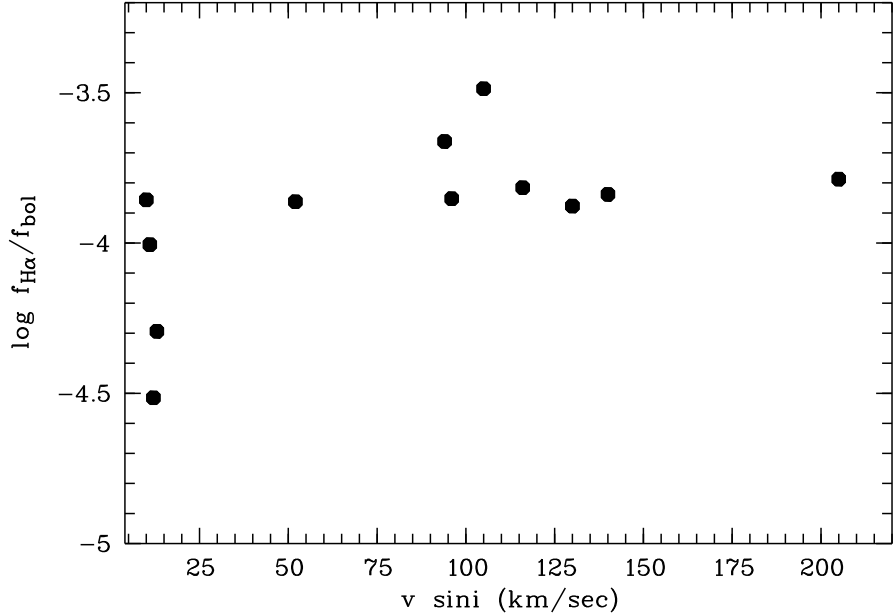}{8cm}{-90}{60}{60}{-230}{320}
\caption{$\log  f_{\rm H\alpha}/f_{\rm bol}$ vs. v$\sin i$ for
$\alpha$ Per.} \label{fig-5}
\end{figure}

In Figure ~\ref{fig-5} $\log f_{\rm H\alpha}/f_{\rm bol}$
is plotted vs. v$\sin i$ for $\alpha$ Per
stars for which H$\alpha$ data are available. 
H$\alpha$ fluxes have been derived from published
equivalent widths using the relationship given by Soderblom et
al. (1993). Although the number of points in the figure is
rather small, the saturation relation seems to be well defined
and there is no evidence for a decline from the saturated ratio
at very high v$\sin i$ values, i.e., supersaturation is not seen for
H$\alpha$. 

I do not have an explanation at hand for supersaturation -probably
we should fully understand saturation first! However, addressing
the problem, and hopefully giving an answer to it, would 
help towards our general understanding of stellar dynamo.
In the following, I shall discuss an observational test which
should permit us a better comprehension of the physical reasons
that lie behind the observed phenomenon of supersaturation.
Namely, two hypothesis can be done. On the one hand, supersaturation
could be the result of an overall decrease of dynamo efficiency
at very high rotation. On the other hand, supersaturation in X-rays
could just mean that more relatively cool plasma (with T $\sim 10^5$
K) is heated, emitting UV and EUV radiation, rather than coronal plasma.
This, on turn, could be a consequence of
the decrease of the apparent surface gravity due to enhanced rotation.
A lower stellar surface gravity implies an increase of the
gravitational scale height of coronal plasma;
more loops
at temperatures of the order of 100,000 K may then exist, since such cool
loops with height lower than the gravitational scale height are stable
(e.g., Antiochos and Noci 1986; Antiochos, Haisch, and Stern 1986).
If this is the case, one would expect an increase of radiative losses
at this temperature, i.e. the emission line flux from the C IV doublet
at $\sim$ 1550 \AA ~should increase. To conclude, C IV fluxes for supersaturated
stars, and in particular the comparison with X-ray data, would give a key
to discern whether supersaturation is indicating 
supersaturation of the dynamo or, more simply, the readjustment of the
radiative output from the corona to lower temperature.

An alternative, completely different explanation which was 
suggested by M. Gagn\`e  during
the discussion at the Workshop (see the question below), is that, with
increasing rotation, coronal temperature also increases (see also A.
Collier Cameron's comment) and most of X-ray emission moves out of
the ROSAT passband. With future X-ray missions
it will indeed be possible to obtain X-ray spectra extending to higher
energies than ROSAT \footnote{ASCA and SAX would not provide a 
high enough
sensitivity for supersaturated UFRs in the $\alpha$ Per or IC2602/2391
clusters.}. This will allow us to
ascertain whether there is a shift of the coronal emission measure
distribution (DEM) outside the ROSAT passband. It is important to note,
however, that McGale et al. (1996) find that W UMa binaries appear to have, in
general, relatively less emitting material at high temperatures than
RS CVn and BY Dra systems.
\section{Summary}

\begin{itemize}
\item ROSAT data for young clusters have shown a decline of the L$_{\rm
X}$/L$_{\rm bol}$ ratio from the saturated value for the so called UFRs.
The decrease starts to be seen at v$\sin i \ga 100$ km/sec, or P$\ga$ 8
hrs, or $\log N_{\rm R} \la -1.8$;
\item supersaturation -that is how this phenomenon has been named- is most
likely a real effect; 
\item it is not clear whether active
RS CVn and BY Dra systems exhibit supersaturation as well; W UMa contact
binaries, instead, show supersaturation;
\item supersaturation is not seen for H$\alpha$ emission, at least as far
as $\alpha$ Per is concerned;
\item whereas supersaturation cannot be easily explained with the current
understanding of stellar dynamo, a straightforward test would allow to
check whether it is the consequence of a decrease of the efficiency
of stellar dynamo or, less dramatically, of the redistribution of the heating
to the cooler, transition region plasma;
\item alternatively, very rapid rotation could lead to a substantially
higher coronal temperature and to the shift of the DEM out of the ROSAT
passband.

\end{itemize}

\acknowledgments

I am grateful to R. Dempsey for sending RS CVn and BY Dra binaries data
and to R. Pallavicini for his useful comments on the manuscript.

\begin{question}{Marc \ Gagn\'e}
(Comment, 1st):
The field M dwarfs and the BY Dra variables show supersaturation as well,
with saturation kicking in at 6 -- 10 km/sec and supersaturation above 40
-- 100 km/sec.

\noindent (Comment, 2nd):
Another interpretation of the ROSAT results is that UFRs have
hotter coronae and that their emission has moved out of the ROSAT
passband.
\end{question}
\begin{answer}{  }
 Yes, I agree on your 2nd comment. Thanks for stressing this.
As to your first comment, it seems to me that it is not actually clear
from the figure I have shown whether BY Dra binaries are supersaturated
or not.
\end{answer}

\begin{question}{Steve Drake}
Were contact binaries included in the plot that you showed of binary
stars' X-ray emission vs. period? I ask because a number of years ago
there were claims that contact binaries transition region and coronal
emission did lie below what was expected from an extrapolation of
(non-contact) RS CVn binaries; this may then be consistent with the claim
of `supersaturation' seen in rapidly rotating cluster stars.
\end{question}
\begin{answer}{ }
 Contact binaries were not originally included in the plot, but added in
afterwards (see Sect. 2) and they indeed appear to be supersaturated.
\end{answer}

\begin{question}{Andrew Collier Cameron}
A couple of comments. First, Vilhu showed that C IV saturates at
relatively low rotation rates, so there is no clear evidence for an
increase
in the 10$^5$ K emission measure from ``supersaturated" stars. Rather,
the coronal temperature increases with increasing rotation rate.

Second, the supersaturation phenomenon appears to set in at rotation rates
such that the co-rotation radius lies within one stellar radius or so of
the surface. The centrifugal forces on the loop plasma should lead to
shrinkage of the coronal volume.
\end{question}

\end{document}